\documentclass[twocolumn,showpacs,aps,pre,
groupedaddress,amssymb,amsmath,nobalancelastpage]{revtex4}

\usepackage{graphicx}
\usepackage{longtable}

\begin{document}

\title{Comparison between step strains and slow steady shear in a bubble raft}
\author{Michael Twardos and Michael Dennin}
\affiliation{Department of Physics and Astronomy, University of
California at Irvine, Irvine, California 92697-4575}

\date{\today}

\begin{abstract}

We report on a comparison between stress relaxations after an
applied step strain and stress relaxations during slow, continuous
strain in a bubble raft. A bubble raft serves as a model
two-dimensional foam and consists of a single layer of bubbles on
a water surface. For both step strains and continuous strain, one
observes periods of stress increase and decrease. Our focus is on
the distribution of stress decreases, or stress drops. The work is
motivated by apparent disagreements between quasistatic
simulations of flowing foam and simulations of continuous strain
for foam. Quasistatic simulations have reported larger average
stress drops than the continuous strain case. Also, there is
evidence in quasistatic simulations for a general divergence of
the average size of the stress drops that only appears to occur in
steady strain near special values of the foam density. In this
work, applied step strains are used as an approximation to
quasistatic simulations. We find general agreement in the
dependence of the average stress drop on rate of strain, but we do
not observe evidence for a divergence of the average stress drop.

\end{abstract}


\maketitle

\section{Introduction}

An open question in the flow of foam is the correspondence between
``true'' quasistatic flow and constant rate of strain in the limit
that the rate of strain approaches zero. (For reviews of foam and
the flow behavior of foam, see for instance
Refs.~\cite{K88,S93,WH99}). Experiments and simulations of model
foams under constant rate of strain clearly exhibit limiting
behavior in which the properties of the system become independent
of the rate of strain for small enough rates of strain
\cite{D95,D97,GD95,DTM01,LTD02,PD03}. This has been referred to as
the {\it quasistatic limit}. However, simulations have also been
carried out in which a small step strain is applied to the system
and the system is allowed to relax to a local energy minimum
\cite{WBHA92,HWB95,ML04}. Such simulations are referred to as {\it
quasistatic}. Surprisingly, results from quasistatic simulations
and results from the quasistatic limit disagree with regard to
certain aspects of the flow. This raises important questions not
only for the flow behavior of foam, but also for a wide class of
complex fluid materials, including granular systems, suspensions,
colloids, and emulsions.

Understanding the quasistatic limit of complex fluids, along with
glasses and supercooled liquids, is important in the context of
the proposal that jamming provides a general theoretical framework
in which to study these systems \cite{LN98,ITPJamming,TPCSW01}.
Jamming refers to the topological crowding of constituent
particles, arresting their further exploration of phase space. The
jamming phase diagram proposes the existence of a ``jammed'' state
of matter as a function of temperature, stress and inverse density
\cite{LN98,TPCSW01}. For materials with a yield stress, such as
foam, there is an important connection between the jamming
transition and the quasistatic limit. One definition of the yield
stress is the value of stress below which a material behaves like
an elastic solid and above which it exhibits ``flow''. A careful
treatment of the yield stress distinguishes between the transition
to plastic deformation and plastic flow. But, for the purposes of
this paper, one can treat the yield stress as the point at which
the material ``unjams''. For materials with a yield stress that
are subjected to a constant rate of strain in the quasistatic
limit, the average stress is essentially the yield stress.
Therefore, these systems exist in a state that is very close to
the jamming transition. So, understanding the behavior of foam, or
other materials, in the quasistatic limit is one way of probing
the nature of the proposed jamming transition.

An open question for the jamming transition is whether or not it
is a ``true'' phase transition. One feature of such transitions is
the existence of divergences that exhibit well-defined scaling
behavior. These issues have been explored in some detail for the
case of zero stress as a function of density \cite{OSLN03}.
However, the question of the behavior at non-zero stress is still
open. For foam, the issue of divergences and scaling behavior has
been explored in some detail, even before the proposal of a
jamming transition. This is particularly true for measurements of
``avalanches'' or stress drops in response to an applied strain.
Under applied strain (whether continuous or step-strain), foam
initially responds in an elastic fashion. In this regime, the
stress increases with strain. For sufficiently large applied
strain, foam undergoes irregular periods of stress increase and
decrease. Loosely speaking, a stress drop is a period of stress
decrease, and stress drops are typically associated with nonlinear
particle rearrangements. A long standing question in the study of
foam is the nature of the distribution of stress drops and the
length scales associated with regions of particle rearrangements.

Stress drops have been studied in a wide range of simulations,
including the bubble model \cite{D95,D97}, the vertex model
\cite{KNN89,KOKN95,OK95}, the q-potts model \cite{JSSAG99}, and a
quasistatic model \cite{WBHA92,HWB95}. (It should be noted that
for periodic foams in two-dimensions, analytic calculations of the
stress under continuous shear have been carried out and
interesting changes in the nature of the stress drops as a
function of the fluid content are predicted \cite{RK90}. However,
these results are not directly applicable to the random systems
discussed here.) Stress relaxations have been measured directly in
experiments utilizing bubble rafts \cite{LTD02, PD03} and
indirectly in other foam systems \cite{GD95,DK97,KE99}. Results
for the distribution of stress drops vary, but essentially divide
into two categories with respect to the nature of stress drops.
For the most part, simulations of a constant rate of strain, even
in the quasistatic limit, report a distribution of stress drops
that has a well-defined average value \cite{D95,D97,TSDKLL99}. In
these simulations, there is no evidence for a diverging length
scale as a function of rate of strain. There is some evidence of
diverging behavior as a function of density, at both the limit of
a completely dry foam \cite{KNN89,KOKN95,OK95} and as a foam melts
\cite{TSDKLL99}. In contrast, a number of quasistatic simulations
\cite{WBHA92,HWB95,ML04}, as well as recent work that models
plastic flow in general \cite{PALB04}, suggest that a diverging
length scale does exist. Experiments that measure stress directly
agree with the constant rate of strain simulations
\cite{LTD02,PD03}. There have been experiments that only measure
bubble rearrangements. These appear to divide along the line of
the simulations, with continuous strain experiments showing no
evidence of a divergence \cite{GD95,DK97} and quasi-static
measurements suggesting the existence of large scale events
\cite{KE99}. These results raise two important questions. Is there
a fundamental difference between quasistatic step strains and
constant rate of strain? Or, is the difference in results simply a
manifestation of differing definitions of stress drops?

This paper compares measurements of stress drops using two
different types of applied strain. First, we reproduce earlier
results for constant rate of strain experiments in bubble rafts
\cite{LTD02,PD03}. Second, we study stress drops in response to
applied step strains that are well separated by periods of
waiting. With these experiments, we are able to compare the impact
of various definitions of stress drops on measurements of the
average stress drop. Also, we directly compare experimental
studies with quasistatic simulations. As will be discussed in more
detail, the step strains studied here are probably not true
``quasistatic'' steps in the same sense as is used in simulations.
However, they do share many qualitative features with quasistatic
steps, and the results provide some insights into differences
between quasistatic step strains and steady rate of strain.

\section{Experimental Setup}

Our system is a two dimensional foam system referred to as a
bubble raft.  The Couette viscometer used to generate applied
strain and measure the resultant stress is described in detail in
Refs.~\cite{app,PD03}. The basic setup consists of two concentric
``cylinders'' that confine the bubbles in an annular region on the
surface of water. The outer cylinder is a Teflon barrier composed
of 12 segmented pieces. The barrier is able to compress and
expand, so as to adjust the density of the bubble raft. It is also
able to rotate to generate either constant rates of strain or
well-defined step strains. The inner barrier is suspended on a
torsion wire and is free to rotate. By measuring the rotation
angle, the stress generated in the bubble raft is measured. The
creation and characteristics of the bubble raft are discussed in
detail in Ref.~\cite{LTD02, PD03}. Essentially, nitrogen gas is
bubbled through a solution of 80\% water, 15\% glycerine, and 5\%
Miracle Bubbles (Imperial Toy Corp.). The needle size and flow
rate is adjusted to select the bubble distribution. A random size
distribution of bubble radii ranging from $1\ {\rm mm}$ to $5\
{\rm mm}$ is used.

In this work, we focus on the nature of stress drops and the
response to different types of strains applied. In addition, we
consider the importance of the definition of a stress drop. To
achieve these tasks, the focus is on step strain measurements.
Step strains are generated by rotating the outer barrier at a
relatively fast constant angular speed for a relatively short time
period. Then, the outer barrier is held fixed for a selected time
interval. This measurement is designed to parallel quasistatic
simulations of foam in which the system is strained an increment
and then energy is minimized. There are two aspects to a ``true''
quasistatic step that must be considered. First, the step itself
should be small enough that it almost always produces a reversible
deformation. Therefore, stress drops, associated with plastic
deformations, should be rare. We will discuss the degree to which
our system captures this feature later. Second, the system is
relaxed until a minimum energy is found. In the experiments, we do
not have access to direct measurements of the energy. So, we can
not determine at what point the energy of the system has achieved
a minimum after the application of a step strain. Therefore, in
order to facilitate comparison with theory, we systematically
increased the waiting time until the results were independent of
the waiting time. The expectation is that, at least in some
statistical sense, this implies we are usually waiting until the
energy is minimized.

There are three main variables of importance that define the step
strains. They are the angular speed of the strain increment,
$\Omega$, the time for rotation, $t_{rot}$, and the time allowed
for relaxation, $t_{rel}$. One way to consider step strain
measurements is to take the total angular displacement applied in
a step and divide it by the time it took to strain it plus the
time allowed for relaxation. This will provide an effective
rotation rate.

\begin{equation}
\Omega_{eff}=\frac{\Omega t_{rot}}{t_{rot}+t_{rel}}
\end{equation}

It should be noted that because we are applying rapid, small
strains, the bubble motions during the strain are essentially
elastic. After the strain is stopped, the bubbles are either
stationary or undergo nonlinear rearrangements. Therefore, the
conversion of effective rotation rate to a rate of strain is not
meaningful. In the case of constant applied rate of strain, the
average bubble motions throughout most of the system are found to
be consistent with various continuum models for fluids
\cite{LCD04}, so a definition of the rate of strain is possible.
Therefore, for the purposes of this paper, the constant rate of
strain results will also be reported in terms of the angular
rotation rate, $\Omega$, of the outer barrier. It is important to
note that the rate of strain in the Couette geometry is a
monotonic function of $\Omega$ \cite{BAH77}. Therefore,
comparisons of effective rotation rate and actual rotation rate
will provide insight into the connection between the quasistatic
limit of constant strain rate and the step strain experiments. For
a given effective rotation rate, we can probe different time
scales and different dynamics by straining it for longer and
longer times while allowing the system to relax by a
proportionally increasingly long time. One question that can be
considered is how an effective rotation rate compares to the
actual rotation rate at that value. Are the stress drop
distributions similar?  Are the average stress values comparable?

It is important to take note of two distinct regimes of step
strains.  On one side are steps so small (much smaller than a
particle diameter) that they are unlikely to independently induce
a stress drop during the strain.  The small step strains from a
theoretical standpoint are the model experimental method to
investigate the long time scale dynamics of infinitely slow strain
rates.  In theory, such an experimental procedure would allow
microstepping strains which would be held by the solid like
properties of the complex fluid in some jammed configuration. On
the other side is step strains that are large, that namely have
reached steady state flow and average stress values by inducing
several particle rearrangements. As the focus of this paper is
comparison with the quasistatic limit of continuous rates of
strain and with quasistatic simulations, we focus on the case of
small step strains. In the following experiments, we selected a
rotation of 0.01 rad/s for 1 second.  This choice was largely
fixed by the physical limitations of the apparatus. However, it
did correspond to a displacement that is less than a typical
bubble diameter throughout the system. These experiments were all
performed close to the time of creation of the bubble raft, within
the first hour of creation when essentially no bubbles were
observed to pop.

\section{Results}

Figure~1 shows a typical response from a series of small step
strains. On the scale of the plot, each step strain corresponds to
a sudden increase in the stress. The subsequent relaxation occurs
during the waiting period. For the first couple of steps, the
system is clearly behaving as an elastic solid, and the stress
remains constant after a step strain. Furthermore, up until
roughly 250~s, the average stress is increasing linearly with the
applied strain. However, above approximately $0.5\ {\rm dyne/cm}$,
the individual steps begin to exhibit relaxation after the initial
stress increase. Finally, the average value of the stress levels
off after sufficient strain at a value of approximately $1\ {\rm
dyne/cm}$. These last two facts are consistent with observations
from continuous rate of strain experiments that suggest a yield
stress on the order of $0.8\ {\rm dyne/cm}$ \cite{PD03,LCD04}.
Also, the agreement between the average stress for the step
strains and the continuous rate of strain suggest that it is
reasonable to compare the two types of flow.

\begin{figure} [htb]
\begin{center}
\includegraphics[width=8cm]{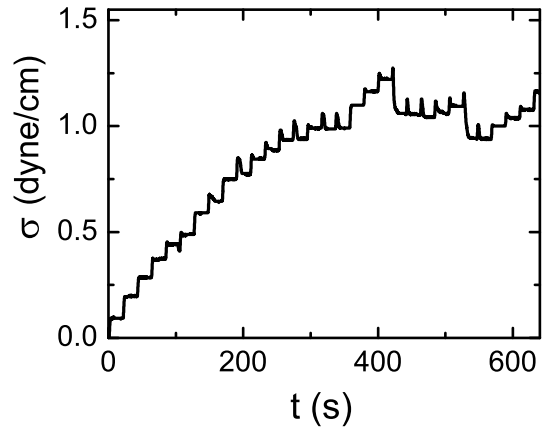}
\caption{A plot of the stress versus time for step strain
measurements with a step time of 1 second at a rotation rate of
0.1 rad/s and a relaxation time of 20 seconds.}
\end{center}
\end{figure}

Figure~2 illustrates a close up of multiple relaxation events that
illustrate the range of responses to a step strain. The dotted
lines mark the end of the step and the onset of a relaxation. Any
given relaxation has two possible outcomes. First, the final
stress can be greater than the stress {\it before} the application
of the step strain. This is a stress increase. Or, the final
stress can be lower than the stress before the application of the
step strain. This is one definition of a {\it stress drop} for the
quasistatic case. Independent of the final value of the stress,
the relaxation process often occurs through multiple relaxations
and plateaus indicative of the complex nature of the stress
relaxation. The plateau regions presumably correspond to
``quasibasins'' during which the energy is still decreasing, but
the decrease is releasing essentially no stress, until the system
suddenly finds itself rapidly approaching a new value of stress.
The fact that multiple plateaus occur complicates the
determination of a true stress minimum during relaxation. This is
one reason why we used multiple waiting times.

\begin{figure} [htb]
\begin{center}
\includegraphics[width=8cm]{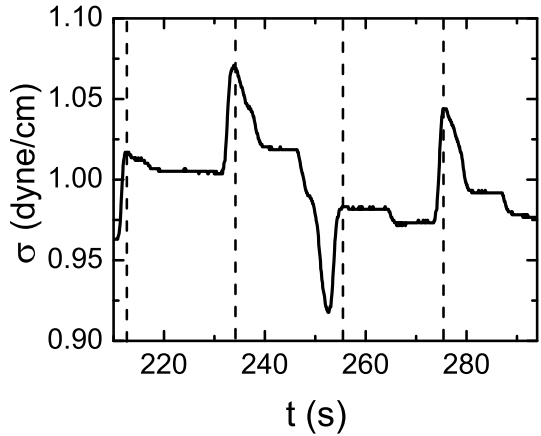}
\caption{A closer view of the stress vs. time graph for step
strains shown in Fig.~1.  The dotted lines mark the end of the
step and the onset of a relaxation.  The multiple relaxations and
plateaus that can be seen following a step strain indicate a
complex relaxation time scale.}
\end{center}
\end{figure}

It should be noted that for waiting times greater than $10\ {\rm
s}$ the relaxation plateaus for at least a few seconds before the
next step is applied in over 90\% of the steps. However, there are
rare events where a stress drop is interrupted. The events in
Fig.~2 were selected to illustrate one such rare event: the step
that occurs at approximately $250\ {\rm s}$.

As mentioned, a stress drop is defined as the difference between
one final value of stress just before a new step strain is made
(i) and the next one (i+1). For purposes of comparing to previous
work, we normalize the change in stress by the average stress for
the given run ($<\sigma>$):
\begin{equation}
\Delta\sigma=(\sigma_{i}-\sigma_{i+1})/<\sigma>
\end{equation}
Because we are mainly interested in the stress drops, it should be
noted that with this definition a stress {\it drop} is {\it
positive}.

As described previously, we report results for an applied rotation
of 0.01 rad/sec for 1 sec and vary the waiting time. Figure~3 is a
plot of the probability distribution for the stress drops for
three different waiting times [20~s ($\circ$), 10~s
($\blacksquare$), and 1~s ($\blacktriangle$)]. (Recall, negative
stress drops are stress increases.) The important feature to note
on the change in stress distribution plots is that the large
stress drop tail increases for increasing waiting times up to the
time scale of about 20~s.  As we will show, for waiting times
greater than 20~s, the tail of the stress drop distribution
appears to be independent of the waiting time. Two other features
of the distribution should be noted. First, the average of the
change in stress (including drops and increases) is essentially
zero. This is important because it implies that a steady-state has
been achieved. The distributions are also asymmetric, with a
longer tail for the stress drops. Therefore, the most probable
event is a stress increase.

\begin{figure} [htb]
\begin{center}
\includegraphics[width=8cm]{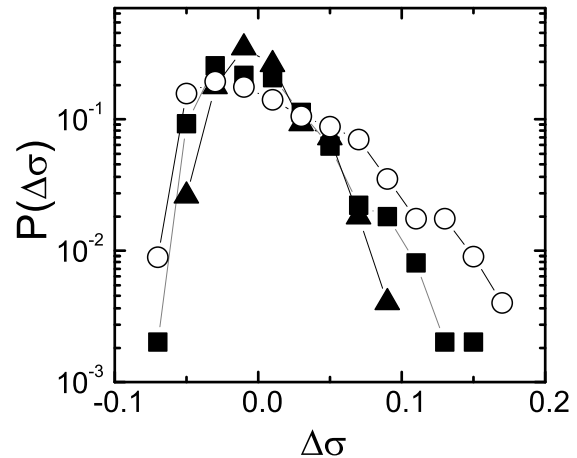}
\caption{A distribution of the change in stress defined as the
difference between one step strain value and the next.  Negative
changes in stress correspond to stress increases while positive
ones correspond to stress drops.  The triangles are a waiting time
of 1 second, the squares are 10 seconds and the circles are 20
seconds. As the waiting time is increased across this regime the
stress distribution broadens and becomes more asymmetric.}
\end{center}
\end{figure}

Consideration of Figs.~2 and 3 in some detail provide insight into
the question raised earlier concerning the true quasistatic nature
of the stress drop. First, Fig.~3 confirms that the most probable
event is a stress increase. However, stress increases represent
only 57\% of the events, not the 90+\% of the events one would
expect in a true quasistatic situation. Also, essentially all of
the stress ``increases'' involve some relaxation of the stress
generated during the applied step strain. This could be due to two
effects. First, the applied step occurs at a finite strain rate.
Therefore, some fraction of the stress increase is due to viscous
effects, and this is expected to relax after the step strain is
complete. Such relaxation does not necessarily involve plastic
events. The other option is that some irreversible events do occur
even during a stress increase. To resolve this issue, detailed
measurements of the bubble motions are required. Therefore, it is
important to keep in mind that the results reported here are for
step strains that only {\it approximate} a quasistatic step.

Figure~4 is a plot of the probability distribution of only the
stress drops and provides a comparison between steady rotation and
step strains for one system size. The number of bubbles was $1.05
\times 10 ^4$. It is important to realize that for continuous rate
of strain, the definition of a stress drop is slightly different.
In this case, because there is no well defined waiting time, a
stress drop is defined as {\it any} decrease in the stress. This
definition was used in our previous measurements
\cite{LTD02,PD03}. We will discuss the implications of this for
the step strain experiments when we discuss the average stress
drop size. The solid symbols are for the three different waiting
times [1~s ($\blacktriangle$), 20~s ($\blacksquare$), and 60~s
($\blacksquare$)]. The open symbols are for the two different
continuous rotation rates [$\Omega = 0.005\ {\rm rad/s}$
($\square$) and $\Omega = 0.002\ {\rm rad/s}$ ($\circ$)].  For
long enough waiting times ($>10\ {\rm s}$), the overall shapes of
the distributions for continuous and step strain measurements are
similar. There is a clear cutoff at large stress drops, and the
probabilities for large stresses are similar. What is not obvious
from this plot is the difference for the small stress drops.
Closer inspection shows that the continuous strain has
significantly more small stress drops. This is not surprising
given the two different operational definitions. For the step
strain case, the entire relaxation is used, even if it is composed
of multiple small steps. The degree to which the small stress
drops dominate the continuous rate of strain is best illustrated
by considering the average stress drop.

\begin{figure} [htb]
\begin{center}
\includegraphics[width=8cm]{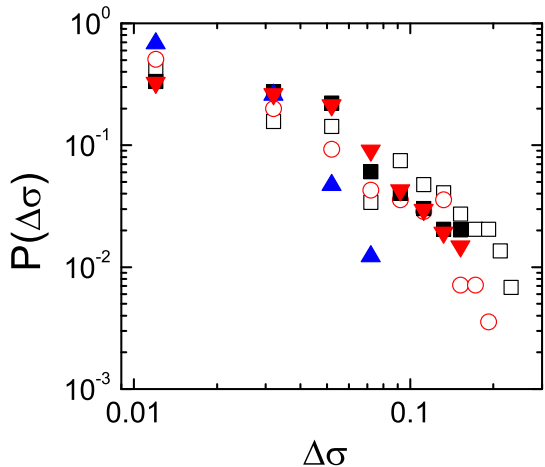}
\caption{Probability distribution for only the stress drops for
both a series of step strains (closed symbols) and for continuous
strain (open symbols). For long enough waiting times (>10 s) the
overall shape of the distribution between the two types of flows
are similar. The three waiting times for the step strain
experiments are 1~s ($\blacktriangle$), 20~s ($\blacksquare$), and
60~s ($\blacksquare$). The two continuous rotation rates are
$\Omega = 0.005\ {\rm rad/s}$ ($\square$) and $\Omega = 0.002\
{\rm rad/s}$ ($\circ$). The results highlight the similarities of
the step strain and continuous strain distributions for large
stress drops.}
\end{center}
\end{figure}

Figure~5 shows the average stress drop as a function of effective
rotation rate for step strains ($\blacksquare$) and actual
rotation rate for the continuous rate of strain ($\blacktriangle$)
measurements. Here the dominance of the small stress drops is
apparent. For the continuous strain case, we have reproduced the
results reported in Ref.~\cite{PD03} that the average stress drop
decreases with decreasing rotation rate. For the step strain case,
we observe the behavior reported for simulations in
Ref.~\cite{PALB04} that the average stress drop increases with
decreasing rate of strain and reaches a well defined plateau.

\begin{figure} [htb]
\begin{center}
\includegraphics[width=8cm]{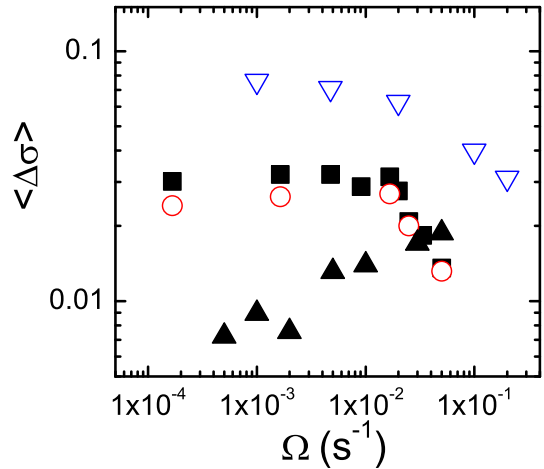}
\caption{Plot of the average stress drop as a function of
$\Omega_{eff}$ for the step strain experiments ($\blacksquare$)
and as a function of $\Omega$ for the continuous strain
experiments ($\blacktriangle$). Also shown are the results for the
step strain experiments with an alternative definition of a stress
drop that counts each individual decrease during an entire
relaxation event ($\circ$), and the alternative definition of
stress drops for continuous strain, as discussed in the text
($\nabla$). The results highlight the differences between the two
types of strain for the average stress drop measurements, as well
as the impact of different definitions of a stress drop.}
\end{center}
\end{figure}

To understand better the impact of the definition of the stress
drop, we can plot two other quantities. First, for the step strain
experiments, we can use the same definition as was used for the
continuous rate of strain experiments, where any period of stress
decrease is taken as a stress drop. This results in an increase in
the number of small stress drops. The results for the average
stress drop in this case are given by the open circles in Fig.~5.
Here we see that this definition does decrease the average stress
drop, but not to the degree that is observed in the continuous
rate of strain case.

We also analyze the stress drops in the continuous rate of strain
case by the method described in Ref.~\cite{PALB04}. In this case,
a stress drop is defined by taking an appropriate time interval,
$\tau$, and computing $\sigma(t) - \sigma(t+\tau)$. The time
interval $\tau$ has to be sufficiently large so as not to
artificially break up a ``typical'' stress drop. This is achieved
by measuring the average stress drop with increasing values of
$\tau$ until the measurement is independent of $\tau$. The average
stress drop is then defined as $<\Delta \sigma> = <\sigma(t) -
\sigma(t+\tau)>/<\sigma>$. Two examples of the dependence of
$<\Delta \sigma>$ as a function of $\tau$ are illustrated in
Fig.~6 for two rotation rates. A number of features of the
behavior are interesting. First, the value of $\tau$ at which
$<\Delta \sigma>$ becomes independent of $\tau$ is an indication
of the typical time over which an event occurs. One can see that
this is of the order of $10\ {\rm s}$ in both cases. Second, the
time appears to scale with rotation rate (or strain rate),
suggesting that the events are best characterized by a typical
strain interval.

\begin{figure} [htb]
\begin{center}
\includegraphics[width=8cm]{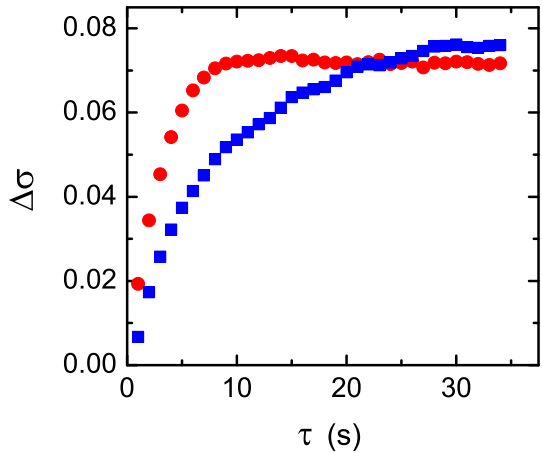}
\caption{Plot of $<\sigma(t) - \sigma(t+\tau)>$ versus $\tau$ for
two different rotation rates. The circles are $\Omega = 0.005\
{\rm rad/s}$, and the squares are $\Omega = 0.001\ {\rm rad/s}$.
In both cases, the curves plateaus for $\tau$ on the order of $10\
{\rm s}$, with the exact value of the plateau proportional to the
rotation rate.}
\end{center}
\end{figure}

The results for $<\Delta \sigma>$ as measured by computing
$\sigma(t) - \sigma(t+\tau)$ are plotted in Fig.~5 ($\nabla$), as
well. Here we again recover the behavior reported in
Ref.~\cite{PALB04} that the average stress drop increases with
decreasing rate of strain. This result is not too surprising
because this alternative definition deemphasizes small stress
drops that occur on short time scales. Also, this definition of
the average stress drop is related to the the variance of the
stress as a function of time. The variance was reported in
Ref.~\cite{PD03}, and behavior similar to that reported in
Ref.~\cite{PALB04} was observed.

\begin{figure} [htb]
\begin{center}
\includegraphics[width=8cm]{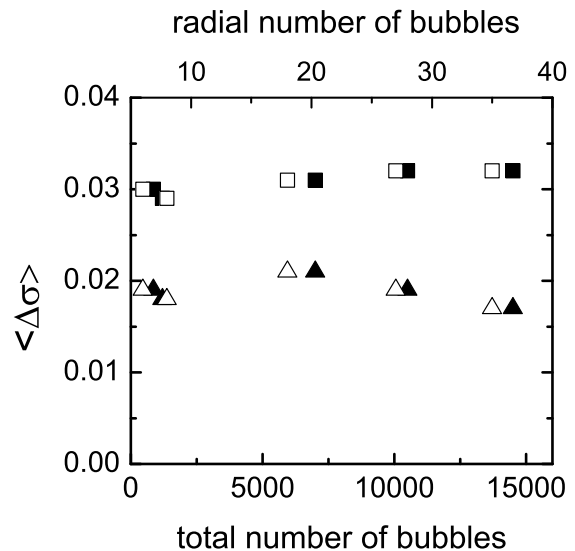}
\vskip 0.2in
\caption{Plotted here is the system size dependence of
the average stress drop for two different waiting times. The
squares are for a waiting time of 60~s, and the triangles are for
a waiting time of 2~s. The open symbols correspond to the top
axis, which gives the average number of bubbles in the radial
direction. The closed symbols are relative to the bottom axis. In
all cases, no observable system size dependence was measured.}
\end{center}
\end{figure}

Finally, we considered the system size dependence of the average
stress drop. This is illustrated in Fig.~7 for four different
system sizes. The system sizes are given using two different
measures: the average number of bubbles in the radial direction
and the total number of bubbles. Also, we show data for two
different effective rotation rates. The triangles correspond to
data with a waiting time of 2~s and the squares correspond to data
with a waiting time of 60~s. In all cases, there is no evidence of
any system size dependence.

\section{Discussion}

In this study, we report on measurements of stress relaxations in
response to small step strain increments followed by a fixed
waiting time. These results are compared with similar measurements
under the application of a constant rate of strain. The results
show interesting connections between step strain measurements and
the constant rate of strain data.

First, the qualitative features of the two measurements are
similar. The distribution of stress changes is asymmetric, with a
tail for large stress drops. There is a well defined average
stress drop for all cases of interest; however, it depends on the
definition of stress drop. For continuous rotation, the simplest
definition of stress drop results in a decrease in average stress
drop with rotation rate. In contrast, a definition that attempts
to capture the concept of an ``event'' rather than an individual
stress drop results in a measured increase in the average stress
drop with rotation rate. This reflects the occurrence of small
stress drops during a single ``stress release'' event.

Another consistent feature of all of the measurements is the
evidence for a ``typical'' time scale for a stress release event.
First, the step-strain measurements suggest a time scale on the
order of $10\ {\rm s}$ based on when these measurements become
independent of the waiting time. The measurements from continuous
rotation that are based on $\sigma(t) - \sigma(t+\tau)$ also
suggest a fundamental time scale on the order of $10\ {\rm s}$. In
this case, it is important to note that the strain rate does enter
as well into determining the average time for an event. Finally,
the previous measurements gave a strain rate of $0.07\ {\rm
s^{-1}}$ for the crossover to a quasistatic limit under continuous
strain \cite{PD03}. This also suggests a fundamental time scale on
the order of $10\ {\rm s}$ for this system. The obvious question
to ask is the source of the time scale. This is easily done within
the context of the bubble model \cite{D95}. In the bubble model,
bubbles are treated as overlapping circles that interact through a
spring force and viscous dissipation. There is only a single time
scale in this model, $\tau \approx b/\sigma$, where $b$ sets the
scale of the viscous dissipation and $\sigma$ is the surface
tension of the bubbles. Future work will involve varying these
quantities to further explore this fundamental time scale.

When comparing the continuous rotation and the step strain
experiments, it is interesting to note that similar definitions of
stress drops (considering  individual decreases in stress between
periods of increase or plateaus) do not give similar results. As
expected, for the step-strain case, this definition results in a
reduction of the measure average stress drop compared to
considering simply the entire stress relaxation during a step. In
contrast, during continuous rotation, this definition results in
an even lower measured average stress drop. This is evidence for
dynamical differences between the slow, steady strain rates and
the step strains that is probably consistent with the results for
the measurement of $\tau$ when determining $<\sigma(t) -
\sigma(t+\tau)>$. Even though $\tau$ is of the same order of
magnitude as other measured time scales, it is clearly dependent
on $\Omega$. As the rotation rate is decreased, the time for a
single event increases. This presumably increases the opportunity
for small stress drops to occur, reducing the measured average
stress drop. In contrast, for step strains, waiting longer does
not impact the number of small stress drops that occur, it simply
increases the length of the final plateau on average. These
results suggest that there is a fundamental difference between a
step strain and continuous rotation that needs to be accounted for
when comparing simulations and experiments. Future studies of the
bubble motions will help elucidate the differences between these
two types of applied strain.

Finally, we have considered the possibility that the average
stress drop is system size dependent. Scaling with system size has
been observed in at least two different simulations of
plastic-type flow \cite{ML04,PALB04}. For the range of system
sizes studied here, we observed no dependence on the system size
for either long or short waiting times in the step-strain
experiments. This was also the case for the average stress drops
measured previously \cite{PD03}. In making this comparison, it
should be noted that the step strains are not truly quasistatic
and the geometry of the experiments is different from the
simulations in a potentially significant fashion. Despite the
clear differences between the continuous strain and the step
strain, it may be that the step strain is still not a sufficiently
good approximation of quasistatic to capture the length scale
divergence.

In regards to the geometry, the simulation uses a square box that
is scaled up in size. For our system, the system size was varied
by increasing the radial dimension of the system. Though this does
lead to a corresponding increase in the azimuthal direction, the
azimuthal direction remains periodic for all system sizes, and the
radial direction has fixed boundaries. This has the effect that
bubble rearrangements in the azimuthal direction are free to be as
large as they want. As we are measuring the azimuthal stress, this
may be the source of the size independence of the average stress
drop. This is an aspect of the experiments that can be explored in
the simulations. Also, experiments are planned to directly measure
the spatial distribution of bubbles involved in the stress
releases. Initial measurements of individual bubble motions were
inconclusive with regard to the issue of the existence of
system-wide events \cite{D04}, so further work is required.

\begin{acknowledgments}

This work was supported by the Department of Energy grant
DE-FG02-03ED46071. The authors thank Craig Maloney, Andrew
Kraynik, and Lyd\'{e}ric Bocquet for useful discussions.

\end{acknowledgments}


\end{document}